\documentclass[a4paper,fleqn]{cas-dc}

\usepackage[authoryear,longnamesfirst]{natbib}

\usepackage{amsmath,amssymb,amsfonts}
\usepackage{algorithmic}
\usepackage{graphicx}
\usepackage{subfigure} 
\usepackage{textcomp}
\usepackage{xcolor}
\usepackage{tcolorbox}
\usepackage{float}
\usepackage{makecell}
\usepackage{colortbl}
\usepackage{diagbox}
\usepackage{multirow}
\usepackage{booktabs}
\usepackage{mdframed}
\usepackage{utfsym}

\newmdenv[skipabove=2mm]{findingbox}

\newcommand{\rerevise}[1]{{#1}}

\def\tsc#1{\csdef{#1}{\textsc{\lowercase{#1}}\xspace}}
\tsc{WGM}
\tsc{QE}

\begin{document}
\let\WriteBookmarks\relax
\def\floatpagepagefraction{1}
\def\textpagefraction{.001}

\shorttitle{Research Artifacts in Software Engineering Publications}    

\shortauthors{Liu et. al}  

\title [mode = title]{Research Artifacts in Software Engineering Publications: Status and Trends}  



%

\author[1]{Mugeng Liu}
\ead{lmg@pku.edu.cn}




\author[1]{Xiaolong Huang}
\ead{huangxiaolong@pku.edu.cn}

\author[1]{Wei He}
\ead{weihe@stu.pku.edu.cn}

\author[2]{Yibing Xie}
\ead{xybybing@stu.pku.edu.cn}

\author[3]{Jie M. Zhang}
\ead{jie.zhang@kcl.ac.uk}

\author[2,4]{Xiang Jing}
\ead{jingxiang@pku.edu.cn}

\author[5]{Zhenpeng Chen}
\ead{zhenpeng.chen@ntu.edu.sg}


\author[6]{Yun Ma}
\cormark[1]
\ead{mayun@pku.edu.cn}

\affiliation[1]{
    organization={Key Laboratory of High Confidence Software Technologies (Peking University), Ministry of Education, School of Computer Science, Peking University},
    city={Beijing},
    country={China}
}

\affiliation[2]{
organization={School of Software and Microelectronics, Peking University},
city={Beijing},
country={China}
}

\affiliation[3]{
    organization={Department of Informatics, King's College London},
    city={London},
    country={ United Kingdom}
}

\affiliation[4]{
    organization={National Key Laboratory of Data Space Technology and System},
    city={Beijing},
    country={China}
}

\affiliation[5]{
    organization={School of Computer Science and Engineering, Nanyang Technological University},
    country={Singapore}
}

\affiliation[6]{
organization={Institute for Artificial Intelligence, Peking University},
city={Beijing},
country={China}
}

\cortext[1]{Corresponding author.}



\begin{abstract}
The Software Engineering (SE) community has been embracing the open science policy and encouraging researchers to disclose artifacts in their publications. 
However, the status and trends of artifact practice and quality remain unclear, lacking insights on further improvement.
In this paper, we present an empirical study to characterize the research artifacts in SE publications.
Specifically, we manually collect 1,487 artifacts from all 2,196 papers published in top-tier SE conferences (ASE, FSE, ICSE, and ISSTA) from 2017 to 2022.
We investigate the common practices (e.g., URL location and format, storage websites), maintenance activities (e.g., last update time and URL validity), popularity (e.g., the number of stars on GitHub and characteristics), and quality (e.g., documentation and code smell) of these artifacts.
Based on our analysis, we reveal a rise in publications providing artifacts. The usage of Zenodo for sharing artifacts has significantly increased. 
However, artifacts stored in GitHub tend to receive few stars, indicating a limited influence on real-world SE applications.
We summarize the results and provide suggestions to different stakeholders in conjunction with current guidelines.
\end{abstract}



\begin{keywords}
 Research artifact\sep Empirical study\sep Software engineering\sep  Code smell
\end{keywords}

\maketitle

\section{Introduction}

Artifacts of publications play a vital role in research\footnote{In this paper, following previous work (\cite{ChristopherTimperley2021}),
we define an artifact as ``any external materials or information provided in conjunction with a research paper via a link''. }.
Artifacts improve the understanding and provide supportive evidence for the claims in a research paper.
They also facilitate code and data reuse and allow for future extension, improvement, and comparison (\cite{ChristopherTimperley2021}). 
Artifacts have gotten more and more attention from the research community since 2016, after a crisis of reproducibility was made open to the public by a Nature’s survey in which more than 1,500 researchers revealed having trouble reproducing previous research results (\cite{baker20161}).

\begin{figure}[!tp]
\begin{center}
\includegraphics[width=0.48\textwidth]{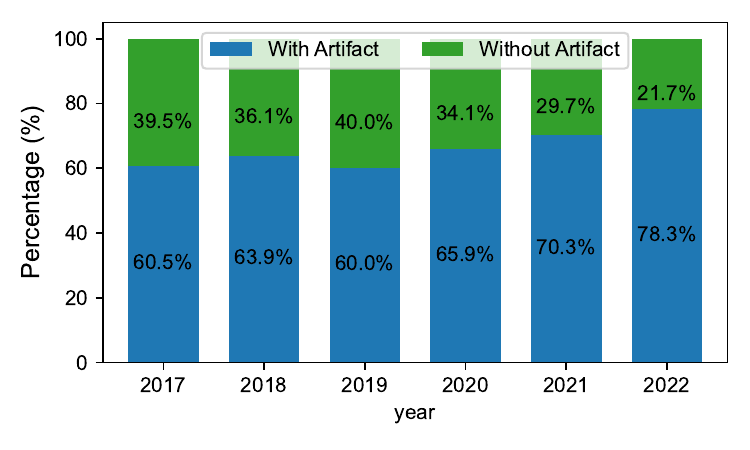}
\end{center}
\vspace{-5mm}
\caption{Number and ratio of top-tier publications with and without artifacts from 2017 to 2022.}
\label{fig:ratioofpaper}
\end{figure}

Software Engineering (SE) community is one of the research communities that have put substantial effort in encouraging open-source artifacts.
Artifact evaluation has become a regular process for software engineering conferences (\cite{hermannCommunityExpectationsArtifacts2020}).
Reproducibility and transparency with artifacts have become one of the key review criteria for top-tier conferences such as 
\cite{icsereviewguideline} 
and 
\cite{fsecfp}.
These efforts have yielded a pleasing increase in the prevalence of open-source artifacts.
As shown in Figure~\ref{fig:ratioofpaper},
the ratio of top-tier publications (ICSE, FSE, ASE, and ISSTA) with artifacts has increased from 60.1\% to 81.1\% from 2017 to 2022 according to our analysis.

Given the importance of artifacts in SE community, some pioneering efforts have been made.
\cite{hermannCommunityExpectationsArtifacts2020} pointed out the community expectations for artifacts by collecting comments from program committee members. 
\cite{ChristopherTimperley2021} explored the general situation of the artifact community from 2014 to 2018 and raised some notable recommendations. 
\cite{krishnamurthi2015real} not only emphasized the importance of repeatability and artifact evaluation but also regarded repeatability as \emph{the real software crisis}.

Despite the efforts from conference organizers and the existing research on the expectations of artifact evaluation, it is unclear how well the current practice of research artifacts performs.
Investigation on the research artifacts in SE publications can be helpful to 
understand the strengths and weaknesses of artifact preparation and be open to a range of measures to improve them.

To this end, this paper presents an empirical study on the status and trends of research artifacts in software engineering publications in the past six years (from 2017 to 2022).
We study the data from 2017 onwards due to the reproducibility crisis that emerged in 2016 (\cite{baker20161}) and its significant impact on the field of software engineering. 
The SE community has witnessed a considerable increase in the concern for the reproducibility of publications since 2017.
Therefore, the data from this time frame can accurately illustrate the current status and trends of SE artifacts.
In particular, we focus on four aspects of artifacts, including 
(1) \textit{common practices}: the types of uploading repositories, programming languages, and the locations in a paper where the links are provided;
(2) \textit{maintenance}: the validity of the provided links and the maintenance time distribution of artifacts;
(3) \textit{popularity}: the star number distribution and the characteristics of top-starred artifacts;
(4) \textit{quality}: the content of documentation and the existence of code smells.

These four aspects illuminate common procedures, durability issues, potential impact, and usability concerns, offering a comprehensive perspective on artifacts within the software engineering community.
Note that we aim to overview the status and trends of SE artifacts in this single paper as much as possible. 
Despite we choose the four most representative aspects, there are still many interesting points worth further study. 
We hope that our work can arouse more interest from the community to work in this direction and provide more systematic and deep studies on real-world artifacts.



To obtain the dataset for analysis, we carefully select the venues to reach the frontier status of artifacts in SE research.
According to the top-tier venue list of SE provided by the \cite{csranking}, we choose the research track of the four software engineering conferences: {International Conference on Automated Software Engineering} (ASE), {ACM SIGSOFT Symposium on the Foundation of Software Engineering/ European Software Engineering Conference} (FSE), {International Conference on Software Engineering} (ICSE), and {International Symposium on Software Testing and Analysis} (ISSTA). 
These conferences are well recognized to be the top-tier four SE conferences (\cite{kim2018precise,kochhar2016practitioners}), and are expected to represent the best practices of research artifacts in the SE community.
We believe that artifacts in these four conferences represent the cutting-edge status and trends in software engineering.

We focus on publications with a minimum of 6 pages, as they are considered to reflect a more mature status compared to shorter papers.
In total, our dataset comprises 2,196 papers. 
We then employ a labor-intensive approach to extract artifact URLs from these publications, resulting in the identification of 1,487 papers offering research artifacts.

Our analysis of these artifacts yields the following primary observations and findings.

In terms of \textit{common practices},
(1) An increasing proportion of SE publications disclosed their artifacts in the past six years (from 60.5\% in 2017 to 78.3\% in 2022), revealing growing recognition of the value of open research artifacts.
(2) Though most SE conferences now explicitly recommend Zenodo over GitHub as an artifact repository platform, the majority (64.2\% in 2022) of researchers choose GitHub for artifacts, likely due to its ease of maintenance and familiarity.
The ratio of Zenodo adoption has increased from 0.0\% to 16.0\% over the past six years. 
(3) Python has overtaken Java to become the most widely used language in artifacts (61.1\% in 2022), calling for more research on quality metrics and automatic examination approaches towards Python, similar to the code smell towards Java.

Regarding \textit{maintenance},
(4) The proportion of link rot\footnote{Due to inactive maintenance, artifacts may become unavailable over time. This phenomenon is called \emph{link rot} (\cite{wwwFetterlyMNW03}), which means that the provided URLs cease to point to their originally targeted artifacts.} raises over time, from 4.8\% in 2022 to 29.8\% in 2017. 
The ratios of link rot on GitHub and dedicated archiving platforms like Zenodo are merely 6.4\% and 7.1\%, suggesting special attention on the long-term availability of artifacts stored outside commonly used artifact service platforms during the review process.
Researchers can leverage tools like the Wayback Machine (\cite{wayback_machine}), robust links (\cite{jones2021robustifying}), or similar technologies to guarantee the enduring availability of their artifacts.
(5) At least half of artifacts require ongoing updates after conferences, highlighting an advantage of GitHub over archiving platforms like Zenodo. Thus, conferences should not blanket recommend against using GitHub but rather should guide researchers to choose the suitable platform for their specific artifacts. 
Additionally, Zenodo can now track updates from GitHub, facilitating collaboration between the two platforms for artifact publication. Researchers may also consider Software Heritage as a durable and citable code archive that considers subsequent commits.

As for \textit{popularity}, 
(6) Most GitHub artifacts receive limited attention, with 65.0\% attracting no more than 10 stars. Though artifact star counts inaccurately gauge popularity, they indicate most software engineering artifacts lack real-world adoption and impact. We believe that rather than only supplement papers, more artifacts should aim to shape real-world applications.

Considering \textit{quality}, 
(7) Over 96\% of projects trigger code smell alerts mainly for code convention rather than functional issues, indicating that code smell detection seems insufficient to accurately assess code quality for artifacts. Research on tailored metrics focused specifically on the code quality of artifacts is necessary.



In summary, this paper makes the following contributions:
\begin{itemize}
    \item We conduct an empirical study on the current status and trends of artifacts in the SE community, which reflects the overall situation.
    \item We present a comprehensive analysis of our empirical results and provide suggestions for different stakeholders to enhance the artifacts practice.
    \item We publicly release our data, scripts, and results for further study.
\end{itemize}

The rest of this paper is organized as follows.
Section~\ref{sec:methodology} describes the methodology of our study. 
Section~\ref{sec:Results} illustrates the experimental results and findings. 
Section~\ref{sec:Discussion} discusses our suggestion and threats to validity. 
Section~\ref{sec:Related Work} describes the related work. 
Section~\ref{sec:Conclusion} concludes the paper. 
Section~\ref{Data Availability} provides our artifact.
\section{Methodology}
\label{sec:methodology}

\begin{figure*}[h!]
\begin{center}
\includegraphics[width=0.65\textwidth]{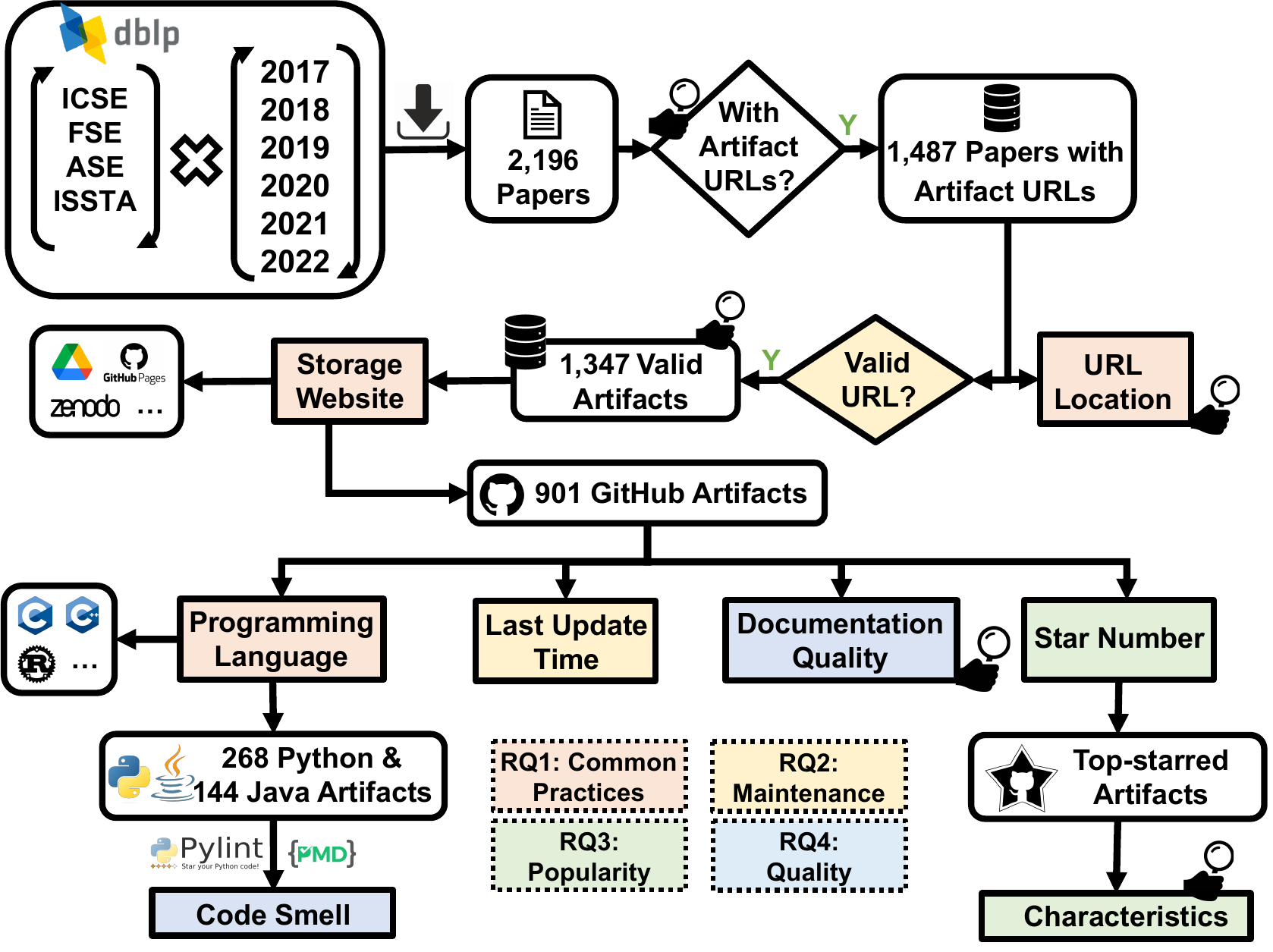}
\end{center}
\caption{
The overview of our methodology.
}
\label{fig:methodology}
\end{figure*}

This section introduces our methodology (Figure~\ref{fig:methodology}), including research questions,
venue selection,
paper and artifact collection process,
and the information extraction details.

\subsection{Research Questions}

To elucidate the status and trends of SE artifacts, we aim to answer the following four research questions (RQs). 

\emph{RQ1. Common practices:} What are the common practices for software engineering researchers to prepare their artifacts? 

\emph{RQ2. Maintenance:} How well do researchers maintain their artifacts?

\emph{RQ3. Popularity:} How is the popularity of existing artifacts and what are the characteristics of top-starred artifacts?

\emph{RQ4. Quality:} How is the quality of the artifacts in terms of the documentation quality and code smells?

Our four RQs encompass four crucial aspects of artifacts, each offering valuable insights into the current state of SE artifacts.

First, RQ1 aims to understand common practices in artifact preparation, serving as a foundation for enhancing the process. By addressing RQ1, we can identify gaps and potential areas for improvement in terms of discoverability, accessibility, and ease of setup.

Second, artifact maintenance directly impacts their longevity and usefulness over time. RQ2 examines the significance of maintenance and investigates existing strategies and potential enhancements.
We specifically focus on link rot, a widely acknowledged phenomenon that significantly undermines the reproducibility of artifacts (\cite{d2015urls,sanderson2011analyzing,klein2014scholarly}).

Third, the popularity of artifacts provides insights into their adoption and usage within the software engineering community, which helps identify artifacts that have made a significant impact and garnered attention from researchers and practitioners.
RQ3 reveals the popularity of current artifacts and provides recommendations for promoting their influence and sharing.

Fourth, ensuring artifact quality is fundamental for reproducibility and reuse. 
RQ4 focuses on documentation (\cite{SoftwareDocumentation2019}) and code smell (\cite{santos2018systematic}) as metrics for quality, which are widely recognized and established measures for evaluating artifacts in the software engineering community.

\rerevise{
On the one hand, the artifact quality heavily depends on the documentation quality, which impacts the clarity, completeness, and ease of reproduction of an artifact.
Clear and well-documented artifacts facilitate reproducibility and allow other researchers to build upon existing work effectively. 
Specifically, a comprehensive ``README'' file serves as the core of documentation that should provide essential information on code, dependencies, file structure, usage, execution examples, etc.
An incomplete ``README'' file can greatly damage the reproducibility of an artifact, even rendering the artifact completely unusable for readers.
}

On the other hand, code plays a vital role in the SE community and makes up a significant part of SE artifacts. 
Code smells can affect the readability, maintainability, and extensibility of code, which ultimately impacts the quality of the research artifacts. 
By analyzing code smells, we can identify potential design issues or violations of good programming practices within artifacts, and contribute to enhancing the code quality of the artifacts.

By addressing these four research questions, we aim to provide a comprehensive understanding of the status and trends of research artifacts in SE.
For ease of presentation, we use Table~\ref{RQs} to demonstrate the content of each RQ.

\begin{table*}[h!]
\caption{Research questions of our study. Each research question is answered from different perspectives.
}
\begin{center}
\resizebox{0.96\textwidth}{!}{\

\begin{tabular}{l|ll}
\toprule
Research Questions & Content & Explanation\\
\midrule
\multirow{3}*{RQ1: Common practices}&Storage websites&What sites or service do researchers use to upload their artifacts?\\
&URL location \&  format&Where in the paper and how do researchers provide their artifacts URL?\\
&Programming language& How does usage of the most popular programming languages vary among artifacts?\\
\midrule
\multirow{2}*{RQ2: Maintenance}
&invalid URLs& How many artifact URLs have become invalid?\\
&Last update time& How does the maintenance of artifacts by researchers relate to conference progress?\\
\midrule
\multirow{2}*{RQ3: Popularity}& Star situation &How many stars do artifacts have?\\
&Characteristics& What are the characteristics of top-starred artifacts?\\
\midrule
\multirow{2}*{RQ4: Quality}&Documentation& How is the quality of the documentation provided by researchers?\\
&Code smell& What are the common code smells in the artifacts?\\

\bottomrule

\end{tabular}}
\label{RQs}
\end{center}
\end{table*}

\subsection{Venue Selection}


To examine the latest advancements in SE artifacts for each year, we focus on venues in the software engineering category of~\cite{csranking}. 
The CS Rankings is a widely acknowledged website that provides an entirely metrics-based ranking system for computer science, which selects the most prestigious publication venues in each area of computer science, including software engineering.
Specifically, we consider ICSE, FSE, ASE, and ISSTA, which are widely recognized as top-tier venues in the field of SE (\cite{kim2018precise,kochhar2016practitioners}), thereby being a good start for the community to understand the best practices in research artifacts.

Overall, we consider these conferences to showcase state-of-the-art research in software engineering, with their associated artifacts capturing the latest advancements in the field each year.

Additionally, we study the data from 2017 onwards due to the reproducibility crisis that emerged in 2016 (\cite{baker20161}) and its significant impact on the field of software engineering. 
The SE community has witnessed a considerable increase in the concern for the reproducibility of publications since 2017.
Consequently, the data from this time frame can accurately illustrate the current status and trends of SE artifacts.

\subsection{Collection of Papers and Artifacts}
\label{sec:method:collection}


To obtain the paper list, we utilize the \cite{dblp} bibliography to search for papers and download them using their DOIs following \cite{khalil2022general}.
DBLP is a widely-used computer science bibliography, ensuring the comprehensiveness of our dataset without omissions.

We further refine our dataset by excluding short papers, following \cite{abou2022software}.
Full papers, which typically present more mature and established results, ensure a more comprehensive representation of the frontier state.
To achieve this criterion, we filter out papers with less than 6 pages, based on DBLP metadata.
 
As a result, we download 2,196 full papers from the research track (i.e., the main track) of four top-tier conferences in the SE community, including ICSE, FSE, ASE, and ISSTA, from 2017 to 2022.

We try our best effort to obtain artifact URLs in papers.
Specifically, we first define a set of search keywords to help us quickly locate the URL, including  ``available'', ``https'', ``replication'', ``reproducibility'', ``code'', ``data'' and some other similar words. 
Then, we conduct a manual search in each paper for artifact URLs by looking for the above keywords. 
If we do not find any URLs during this search, we browse the paper to check if any URLs are mentioned as artifacts again.
Once we have identified these URLs, we manually verify if they correspond to the respective paper.

We improve the accuracy of the annotation results by double-checking.
In detail, two authors of this paper independently annotate the URLs for each paper and cross-check them once all annotations are completed. 
In case of conflicts, we hold discussions to reach a consensus. 

\begin{figure*}[h!]
\begin{center}
\includegraphics[width=0.95\textwidth]{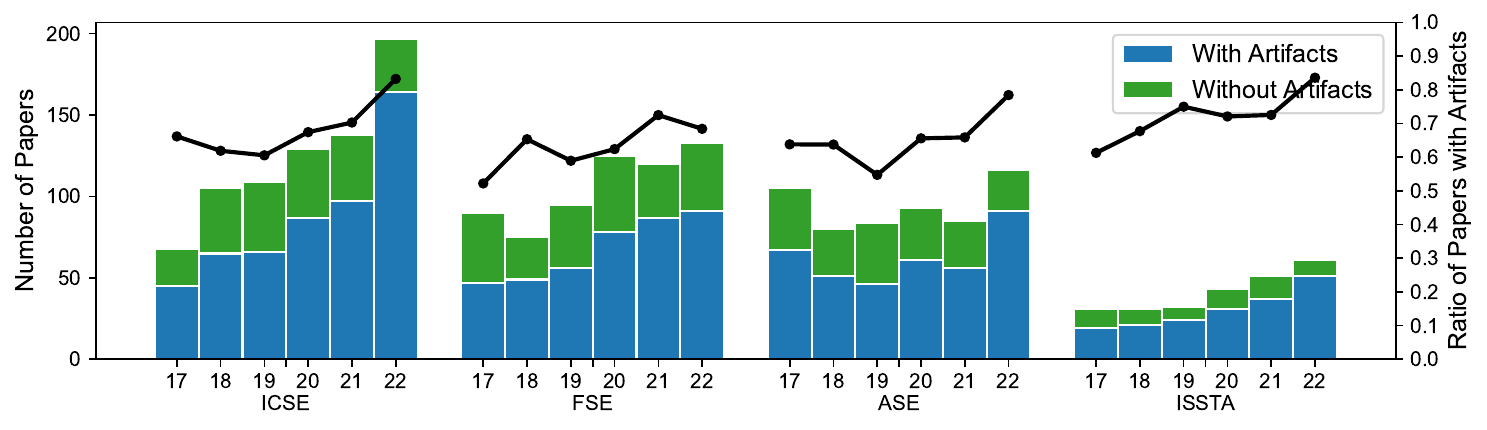}
\end{center}
\vspace{-5mm}
\caption{
Trends of papers with artifacts.
}
\label{fig:trends}
\end{figure*}

Based on our parsing methods, we find 1,487 papers containing artifact URLs.
Table~\ref{tab:detail} shows the results of our collection of papers and artifacts for the four top-tier SE venues from 2017--2022.
Each cell presents the number of papers with artifacts followed by the total number of papers, with the corresponding percentage of papers with artifacts provided in parentheses.
We also present the growing trends of papers with artifacts in Figure~\ref{fig:trends}.

\begin{table*}[h!]
\caption{
Results of papers with artifacts and the total number of papers.
}
\begin{center}
\resizebox{0.98\textwidth}{!}{
\begin{tabular}{lrrrrrrr}
\toprule
Venue & 2017 & 2018 & 2019 & 2020 & 2021 & 2022 & Total \\
\midrule
ASE & 67/105 (63.8\%) & 51/80 (63.7\%) & 46/84 (54.8\%) & 61/93 (65.6\%) & 56/85 (65.9\%) & 91/116 (78.4\%) & 372/563 (66.1\%) \\
FSE & 47/90 (52.2\%) & 49/75 (65.3\%) & 56/95 (58.9\%) & 78/125 (62.4\%) & 87/120 (72.5\%) & 91/133 (68.4\%) & 408/638 (63.9\%) \\
ICSE & 45/68 (66.2\%) & 65/105 (61.9\%) & 66/109 (60.6\%) & 87/129 (67.4\%) & 97/138 (70.3\%) & 164/197 (83.2\%) & 524/746 (70.2\%) \\
ISSTA & 19/31 (61.3\%) & 21/31 (67.7\%) & 24/32 (75.0\%) & 31/43 (72.1\%) & 37/51 (72.5\%) & 51/61 (83.6\%) & 183/249 (73.5\%) \\
Total & 178/294 (60.5\%) & 186/291 (63.9\%) & 192/320 (60.0\%) & 257/390 (65.9\%) & 277/394 (70.3\%) & 397/507 (78.3\%) & 1487/2196 (67.7\%) \\
\bottomrule
\end{tabular}}
\end{center}
\label{tab:detail}
\end{table*}

\subsection{Information Extraction}

To answer the research questions, we extract the following information from collected papers and artifacts.

\subsubsection{Paper Examination}

When obtaining the paper list, we gain the publication year, venue, and title of papers from DBLP.

During the process of manually exploring the artifact URL, we record the location and format of artifact URLs for papers in which we find artifacts.
Typically, we record the URL location and format where we first find the URL in a paper, which can be considered the most prominent placement of the URL.
We categorize the \textbf{URL location} into title, abstract, introduction, conclusion, and other sections because the first four sections are common to all papers.
We also mark the \textbf{URL format} that researchers choose to provide the URL, which is classified into footnote, reference, in-text, and hyperlink. 

\subsubsection{Artifact Examination}

We examine each artifact to collect the information for the content we list in Table~\ref{RQs}.

\noindent\textbf{Storage websites:}
We classify the storage websites of our collected artifacts into GitHub\footnote{\url{https://github.com/}}, Zenodo\footnote{
\url{https://zenodo.org/.} Zenodo is a general-purpose open repository for researchers to deposit any form of research-related digital artifacts. }, self-built homepages (with support from GitHub, Google, or other websites), and others (primarily temporary data drives).
There are 4.6\% of papers utilizing multiple types of storage websites.
We document all the employed storage websites for each paper.

\noindent\textbf{Programming languages:} 
We observe that the storage website of most artifacts is GitHub, which will display the proportion of programming languages used in artifacts.\footnote{Linguist is used on GitHub.com to detect blob languages, ignore binary or vendored files, suppress generated files in diffs, and generate language breakdown graphs. For more information of linguist, check \url{https://github.com/github/linguist} .} 
By leveraging the GitHub APIs, we identify the programming languages utilized by GitHub artifacts, yielding 794 artifacts with the corresponding programming languages.


\noindent\textbf{Last update time:} Utilizing GitHub APIs, we record the latest update time for 901 GitHub artifacts. 
The latest update can be regarded as the most recent modification made to an artifact, specifically the time of the last push on GitHub.

Next, we manually search the web for the important dates of conferences each year, including the paper submission deadline, paper notification time, camera-ready deadline, and conference time.
We record the ratio of artifacts that get updated after each deadline.

\noindent\textbf{Invalid URLs:} 
We record whether the URLs of artifacts provided in papers are valid or not by manually accessing each URL.

\noindent\textbf{Star number:}
``Star'' on GitHub can be described as a form of social endorsement or signaling. 
When starring a repository, the repository will be added to the starred repository list of users, which serves as a personal bookmark and can also indicate to other users that the repository is worth checking out.
As more users star a repository, it becomes more visible within the GitHub community. 
The number of stars a repository has can serve as social proof of its usefulness or interest to potential users, demonstrating that it has been evaluated and endorsed by others.

Therefore, the number of a repository's stars is strongly linked with its overall popularity. In total, we record 783 artifacts with their star numbers through the GitHub API.

\noindent\textbf{Characteristics of top-starred artifacts:}

According to the distribution of stars, only 3.7\% of GitHub artifacts boast more than 100 stars. 
Therefore, we consider these artifacts as top-starred and explore their characteristics. 
Employing this criterion, we identify a total of 33 top-starred artifacts.
In addition to the preceding collected information, we record the documentation quality, executability, number of forks and issues, as well as the latest update time associated with these top-starred artifacts.

\noindent\textbf{Documentation:} 

When examining an artifact on GitHub, we record the documentation quality by manually checking the README file in the artifact. 
Referring to \cite{SoftwareDocumentation2019}, we selected six simple and feasible criteria to measure artifacts. Each criterion is marked by yes or no.

The specific criteria are :
(1) Completeness: Whether the README contains the system version, software, and library environment requirements. \footnote{If the README says ``Please check this file for this information,'' we'll check that file and consider it part of the README. If this file meets the criteria, we also consider the README file to meet the criteria. Unless otherwise stated, the following will follow this principle.}
(2) Structure: Whether the README describes the file structure of the artifact, no matter whether it is accurate to each file or roughly what is in each folder.
(3) Usability: Whether the README contains how to use the artifact.
(4) Example: Whether the README contains any use examples or display of the effect when used.
(5) Certificate: Whether the README states this artifact has a certificate or a certificate can be found along with the README. Furthermore, we record the type of the certificate.
(6) Contact: Whether the README contains the researchers' contact information or the researchers' reply on the topic. \footnote{Notice that on GitHub, the owner of every repository can be found. However, we believe that only the two ways mentioned above reflect the positive willingness of researchers to contact. So these are the only two ways that we would consider satisfying this criterion.} 

Considering the substantial manual effort required for this examination, and recognizing that it is just one aspect of our overall study, we focus specifically on GitHub artifacts from the ICSE spanning the years 2017 to 2022. Given that ICSE features a larger number of papers compared to the other three venues, we consider this subset to be representative enough for our documentation analysis.

\noindent\textbf{Code smells:}

Code that follows a consistent style is much easier to read, which can help others to understand, maintain, and reuse the artifact more efficiently. 
So the code style is crucial to the overall quality of artifacts.

According to the statistics mentioned earlier in ``Programming languages'', Python and Java are the two most used programming languages that dominate the artifacts we collect.
In addition, these languages are supported by well-established code quality metrics (i.e. code smell) and detection tools.
Therefore, we focus on Python and Java artifacts to explore the prevalence of code smell. 

We use \cite{pylint} to collect the code smell information for Python, and \cite{pmd} to collect the information for Java.
We choose these two tools due to their popularity and reliability.
The GitHub repository of Pylint and PMD has over 4.1k stars and 3.9k stars, respectively.
In particular, the Pylint is based on the 
\cite{PEP8}, a style guide for Python code. 


The code smell detection tools can only generate comprehensive and precise code smell reports for artifacts that primarily use ``Python'' or ``Java'' code.
To this end, we use the GitHub API to retrieve the primary programming languages for each repository. 
Therefore, we detect code smells the artifacts that primarily use "Python" or "Java" based on the information provided by the GitHub API. 
To avoid potential infinite loops within a single artifact, we set a time limit of 60 seconds for each detection, which is typically adequate for the majority of artifacts.
As a result, we successfully perform code smell detection on 268 Python artifacts and 144 Java artifacts.


\section{Results}
\label{sec:Results}
This section introduces the results of our study.
For each research question, we present the analysis results and discuss our observations and conclusions.

\subsection{RQ1: Common Practices}
RQ1 aims to understand the common practices of SE researchers when preparing artifacts. To answer RQ1, we classify the collected artifacts according to the categories in each aspect (i.e., storage website, URL location, and programming languages), and 
analyze the distribution of the artifacts.

\subsubsection{Storage Website}

\begin{figure}[!tp]
\begin{center}
\includegraphics[width=0.48\textwidth]{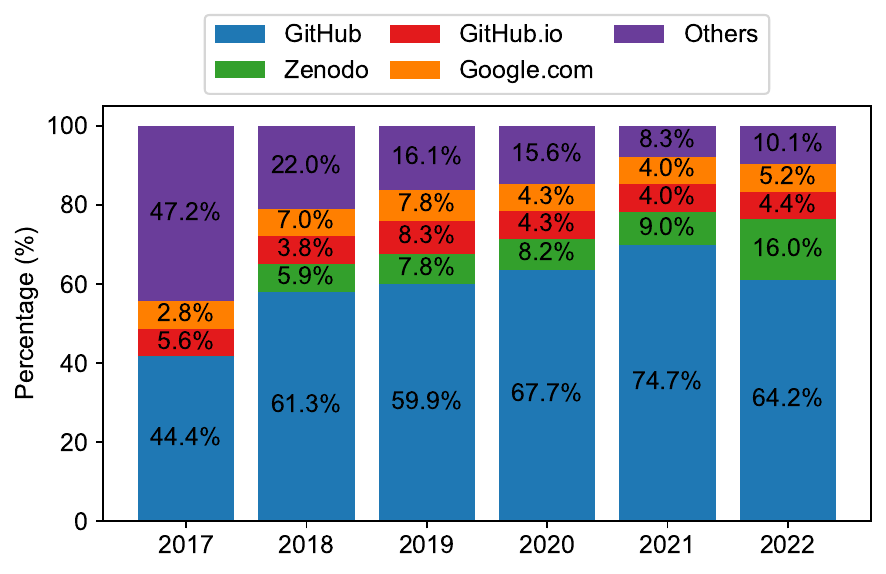}
\end{center}
\vspace{-5mm}
\caption{Distribution and trend of storage website adoption for SE artifacts from 2017 to 2022.}
\label{storagewebsite}
\end{figure}

Figure~\ref{storagewebsite} shows the detailed results of storage websites of artifacts. 
We observe that GitHub is the most widely used website and has gained more and more adoption since 2017.
In 2022, 64.2\% of the artifacts are stored in GitHub repositories.

The second most popular website is Zenodo.
The ratio of artifacts on Zenodo has increased dramatically from 0.0\% to 16.0\% from 2017 to 2022.
Zenodo is a dedicated platform for sharing artifacts, providing researchers with a seamless and anonymous process for uploading their artifacts.
Besides, researchers' operations are recorded on Zenodo. 
When researchers update their artifact, a URL for an artifact on Zenodo will show the original artifact with a hint that there is a newer version, while a URL for an artifact on GitHub will just show the newest version without the availability of the original version.
In addition, every upload in Zenodo is assigned a Digital Object Identifier (DOI), to make the artifact citable and trackable. 
Our observation that Zenodo is getting more adoption aligns with the ``Submission and Reviewing Guidelines'' of the FSE 2021 Artifact Evaluation track (\cite{fsereviewguideline}) (abbreviated as ``FSE21 guideline'' in the remaining part of this paper)
which ``strongly recommends relying on services like Zenodo to archive repositories''.
The open science policy of ICSE 2021 to 2023 also suggests authors archive artifacts on preserved digital repositories such as Zenodo.
As a result, the ratio of artifacts on Zenodo in 2022 (16.0\%) is almost twice as much as it was in 2021 (9.0\%).

The remaining artifacts are published on individual/ institutional websites or temporary drives (e.g., Dropbox, Google Drive, GitLab, and OneDrive). The ratio of these artifacts has decreased from 55.6\% to 19.7\% in the past six years.
The decrease in the ratio of artifacts on individual/institutional websites or temporary drives is also a positive change because these websites are prone to changes and are regarded as non-persistent (\cite{fsereviewguideline}). Besides, These artifacts face more Link Rot as shown in RQ2.2.


Archived on preserved digital repositories such as \url{zenodo.org}, \url{figshare.com}, \url{www.softwareheritage.org}, \url{osf.io}, or institutional repositories such as GitHub, GitLab, and similar services have version management and update time record capabilities. 
Researchers can do vigorous content management on their artifacts under the protection of historical versions without worrying about an error of operation and the coming problems that cause the devastation are difficult or impossible to recover from. For the visitors of the artifact, they can find the specific version or the specific files uploaded at a certain time, instead of just the newest version of the whole artifact. Such advantages are not possessed by Personal or institutional websites, consumer cloud storage such as Dropbox, or services such as \url{Academia.edu} and \url{Researchgate.net}.

\begin{findingbox}{
The majority  (64.2\% in 2022) of publications still upload their artifacts on GitHub, even though services for version control systems are not recommended in some conferences (\cite{icse23policy}).
The ratio of Zenodo adoption has increased \textbf{from 0.0\% to 16.0\%} over the past six years. Due to its strengths and the conferences' recommendations, the usage ratio of Zenodo has been and will continue to grow rapidly.
}\end{findingbox}

\subsubsection{Programming Language}
\begin{figure}[!tp]
\begin{center}
\includegraphics[width=0.45\textwidth]{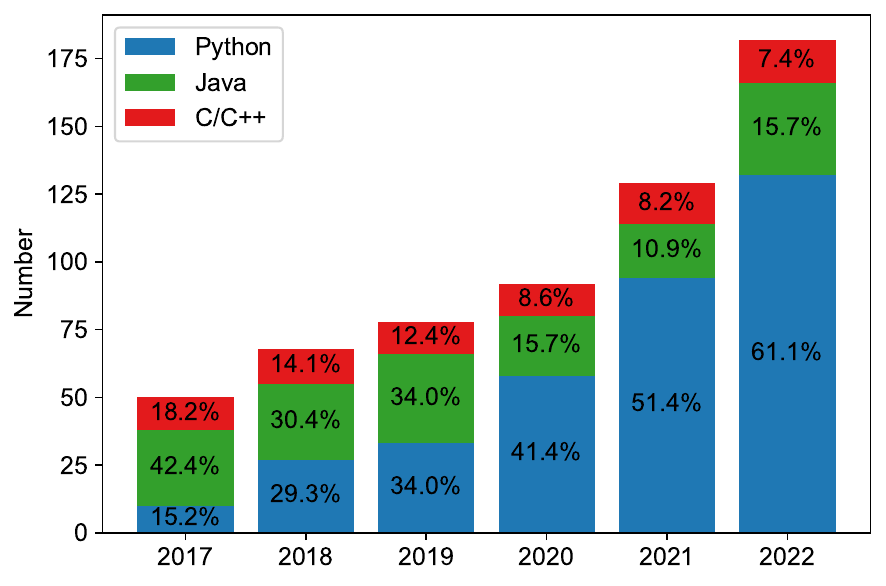}
\end{center}
\vspace{-5mm}
\caption{The number and ratio of artifacts with different programming languages.}
\label{progamminglanguage}
\end{figure}

This part focuses on the programming languages contained in the artifacts. 
The programming language is a core characteristic of most SE artifacts, though not all the artifacts contain code. 
Figure~\ref{progamminglanguage} shows the ratio of artifacts with the three most widely adopted programming languages: Python, Java, and C/C++.

We observe that Python has become more and more widely adopted over the past six years. 
In 2019, Python replaced Java as the most used language.
In 2021 and 2022, over half of the artifacts have Python code.
Java is still the second most widely used language, although its ratio has been decreasing year after year. 
The ratio of C/C++ is also decreasing.

The rising prominence of Python within the computer science community has driven this transformation. The expanding capabilities and user-friendly nature of Python libraries have simplified tasks that were previously accomplished using other programming languages. Consequently, possessing a fundamental understanding of Python has become increasingly essential for researchers.

\begin{findingbox}
Python has overtaken Java and become the most widely used language in SE artifacts and is getting more and more adoption, with its ratio increasing from 15.2\% in 2017 to 61.1\% in 2022. The ratios for Java and C/C++ are both decreasing. 
\end{findingbox}

\subsubsection{URL Location}
During the artifact collection process of our study, we observe that the location of the URL for an artifact in a paper makes a great difference in how easy it is for readers to notice its existence.
For instance, artifact URLs mentioned in the abstract or highlighted in the introduction section are easier to access than URLs given as textual implementation details.

\begin{figure}
\begin{minipage}{0.49\linewidth}
  \centerline{\includegraphics[width=\textwidth]{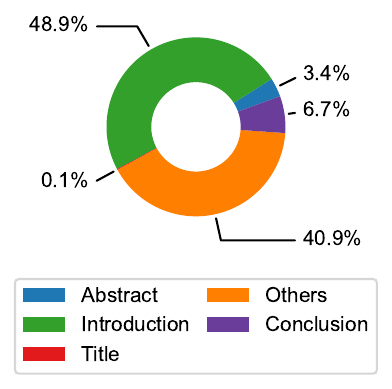}}
  \centerline{(a)}
\end{minipage}
\hfill
\begin{minipage}{0.49\linewidth}
  \centerline{\includegraphics[width=\textwidth]{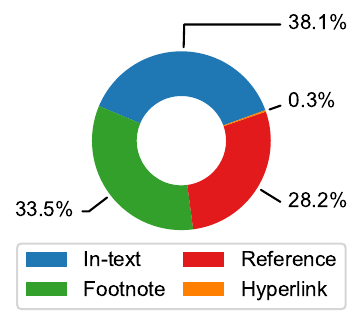}}
  \centerline{(b)}
\end{minipage}
\vfill
\caption{Distribution of artifact URL locations and formats provided in papers.}
\label{fig:location}
\end{figure}

Figure~\ref{fig:location}.(a) shows the distribution of artifact URL locations in different paper sections.\footnote{While some researchers multiply references to the URL, we count them in the first place we find it.}
We observe that 52.3\% of the papers with artifacts highlight the URL of artifacts in the abstract or introduction section.

The open-science policy such as 
\cite{icse23policy} recommends authors provide artifacts in the section of Data Availability after the Conclusion section. 
In our collection, we find that only 5.0\% of all the URLs align with this recommendation in 2017-2021, which are provided with a separate section ``Data Availability'' after the conclusion. 
In 2022, however, this changes greatly. 11.4\% of the URLs are provided as recommended, demonstrating that the policy of ICSE 2023 and other similar recommendations have made a difference. 

40.9\% of the URLs are shown in the section that illustrates implementation details or other similar sections, which are more difficult to notice and search in our practice.

Figure~\ref{fig:location}.(b) shows the format that researchers choose to provide the URL. 
The distribution of URL formats is fairly even, with footnotes, references, and in-text citations each comprising around one-third of the total number of URLs: footnotes (33.5\%), references (28.2\%), and in-text citations (31.7\%).
4 (0.3\%) artifacts are provided with hyperlinks in the text.
Although there is no criterion for researchers to arrange their URL, we recommend not to provide the URL with a hyperlink such as ``click here'', for we find it is quite easy to miss the URL.

\begin{findingbox}
{About half (52.3\%) of the publications with artifacts place the URL of artifacts in the abstract or introduction. Under the recommendation of ICSE 2023 and other conferences, 11.4\% of URLs in 2022 are provided in the section of Data Availability after the conclusion section.}
\end{findingbox}

\subsection{RQ2: Maintenance}
To answer RQ2, we analyze the validity of the URL and the last update time of an artifact.

\subsubsection{Link Rot}
\begin{figure}[!tp]
\begin{center}
\includegraphics[width=0.45\textwidth]{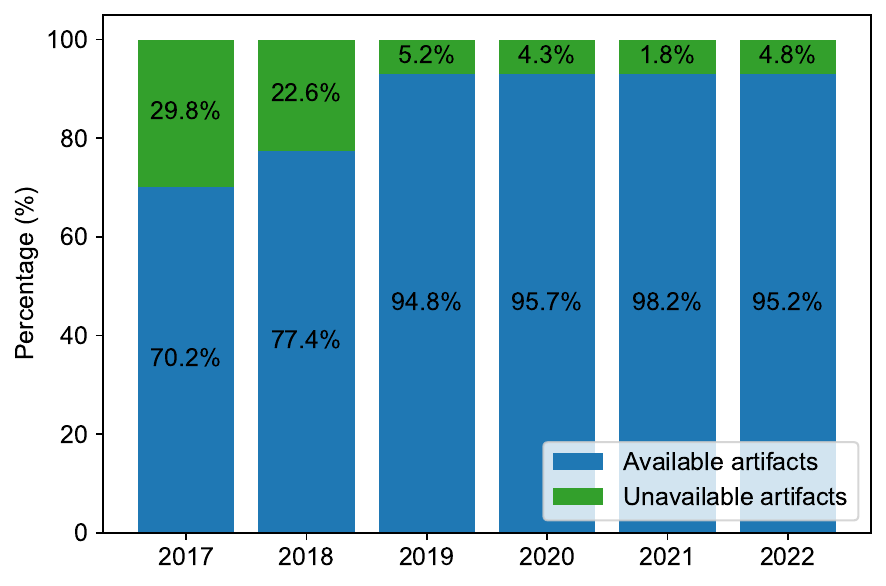}
\caption{The ratio of invalid URLs from 2017 to 2022.}
\label{invalidurls}
\end{center}
\end{figure}

Figure~\ref{invalidurls} illustrates the frequency of link rot in the SE research artifacts from 2017 to 2022. Overall, the proportion of artifacts suffering from the link rot issue increases over time, with an average of 9.4\%. Specifically, the proportions for these years are 29.8\%, 22.6\%, 5.2\%, 4.3\%, 1.8\%, and 4.8\%, respectively. 
Nearly one-third of artifacts provided in 2017 are unavailable due to link rot, around six times the ratio for 2022.

This finding is consistent with that observed by previous work (\cite{ChristopherTimperley2021}). 
In 2019,  \citeauthor{ChristopherTimperley2021} analyzed the URLs extracted from technical papers published in ICSE, FSE, ASE, and EMSE between 2014 and 2018. They found that 26.47\% of artifacts provided in 2014 were unavailable due to link rot, while the corresponding ratio for 2018 is 5.43\%.

\begin{figure}[!tp]
\begin{center}
\includegraphics[width=0.45\textwidth]{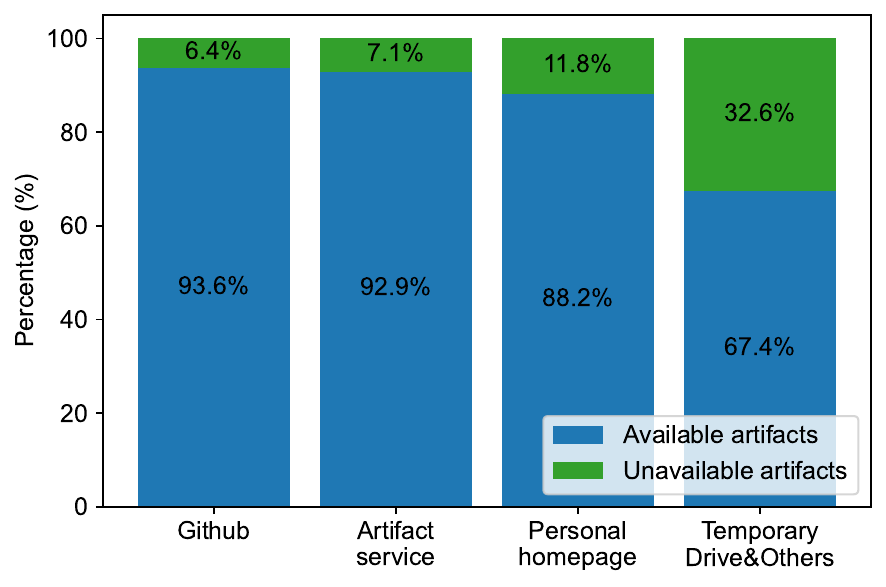}
\caption{The proportions of unavailable artifacts in different types of storage websites.}
\label{invalidurls-which}
\end{center}
\end{figure}

We further analyze the unavailable artifacts in different types of storage websites. Figure~\ref{invalidurls-which} presents the results. We find that 32.6\% of artifacts originally stored on temporary drives (such as Dropbox and Google Drive) have become unavailable. 
Similarly, 11.8\% of artifacts on personal homepages are inaccessible. 
In contrast, the corresponding proportions of artifacts on GitHub and other artifact service platforms (such as Zenodo) are 6.4\% and 7.1\%. 
For the artifacts on GitHub, only when the researchers delete or hide the repository, does link rot happen. Updating does not matter, but for the artifacts on temporary drives, exceeding the sharing time limit or stopping sharing can contribute to link rot. The server changes, domain expiration, and personnel changes can all become the cause of link rot.


\begin{findingbox}
Overall, 9.4\% of the artifacts we collected are unavailable due to link rot. 
The proportion of link rot raises over time, from 4.8\% for publications in 2022 to 29.8\% in 2017. 
Link rot is more prevalent in temporary drives (32.6\%) and personal homepages (11.8\%) than on GitHub (6.4\%) and artifact service platforms (7.1\%). 
We suggest sharing artifacts through artifact service platforms.
\end{findingbox}

\subsubsection{The Latest Update Time}
Table~\ref{tabel:maintaince} shows the percentage of artifacts that were updated after their respective deadlines. 
For example, for the ICSE 2017 conference, 64.3\% of the artifacts were updated after the conference date.

More than 90\% of the artifacts were updated after the submission deadlines each year. 
However, the update rate drops after each subsequent milestone date. 
For instance, for ICSE 2022, 99.0\% of the artifacts were updated after the submission deadline, 91.1\% after the paper notification date, 77.2\% after the camera-ready deadline, and only 54.5\% after the conference date.


When comparing the ratios of the same conference across different years, it is found that the percentage of artifacts updated after the conference date decreases over time. 
For example, in ICSE, 64.3\% of artifacts were updated in 2017, whereas 54.5\% were updated in 2022. A similar pattern is observed for the other three conferences. 
We speculate that older artifacts may receive regular attention and updates to fix bugs over time, while newer artifacts may not have received sufficient attention and encountered problems that require updates.

In general, the update rate of artifacts decreases over time for each conference. This could be due to projects losing maintenance or certain artifacts not requiring regular updates. However, for artifacts that implement tools, the lack of maintenance can make them increasingly difficult to use. It is challenging to measure researchers' level of care for their artifacts. We hope that researchers view artifacts as valuable assets that can be lost if not taken care of, rather than as mere appendages to papers that are forgotten over time.

\begin{table}[h!]
\normalsize
\caption{The maintenance situation of ASE, FSE, ICSE and ISSTA 2017-2022. Each number presents the ratio of artifacts that gets updated after each key time point.}
\begin{center}
\resizebox{0.49\textwidth}{!}{
\begin{tabular}{llrrrr}
\toprule
\multirow{2}{*}{Conference}&\multirow{2}{*}{Year}&Submission&Paper&Camera-ready &Conference\\
&& deadline & notification & deadline &  time\\
\midrule
ASE & 2017 & 92.3\% & 88.5\% & 80.8\% & 76.9\% \\
& 2018 & 96.4\% & 89.3\% & 85.7\% & 78.6\% \\
& 2019 & 96.4\% & 85.7\% & 60.7\% & 53.6\% \\
& 2020 & 100.0\% & 95.6\% & 73.3\% & 62.2\% \\
& 2021 & 100.0\% & 97.8\% & 69.6\% & 52.2\% \\
& 2022 & 98.3\% & 93.2\% & 71.2\% & 55.9\% \\
& Average & 97.8\% & 92.7\% & 72.8\% & 61.2\% \\
\midrule
FSE & 2017 & 100.0\% & 92.0\% & 80.0\% & 68.0\% \\
& 2018 & 96.4\% & 89.3\% & 57.1\% & 46.4\% \\
& 2019 & 97.0\% & 90.9\% & 84.8\% & 78.8\% \\
& 2020 & 95.6\% & 95.6\% & 71.1\% & 60.0\% \\
& 2021 & 98.6\% & 92.9\% & 75.7\% & 47.1\% \\
& 2022 & 96.0\% & 92.0\% & 68.0\% & 46.0\% \\
& Average & 97.2\% & 92.4\% & 72.9\% & 55.4\% \\
\midrule
ICSE & 2017 & 100.0\% & 85.7\% & 85.7\% & 64.3\% \\
& 2018 & 97.1\% & 97.1\% & 85.3\% & 76.5\% \\
& 2019 & 100.0\% & 94.9\% & 84.6\% & 64.1\% \\
& 2020 & 98.1\% & 92.3\% & 73.1\% & 50.0\% \\
& 2021 & 96.7\% & 95.1\% & 78.7\% & 60.7\% \\
& 2022 & 99.0\% & 91.1\% & 77.2\% & 54.5\% \\
& Average & 98.3\% & 93.0\% & 79.1\% & 59.1\% \\
\midrule
ISSTA & 2017 & 100.0\% & 100.0\% & 100.0\% & 100.0\% \\
& 2018 & 100.0\% & 84.6\% & 84.6\% & 76.9\% \\
& 2019 & 90.9\% & 90.9\% & 81.8\% & 63.6\% \\
& 2020 & 95.8\% & 95.8\% & 83.3\% & 70.8\% \\
& 2021 & 96.4\% & 92.9\% & 82.1\% & 60.7\% \\
& 2022 & 100.0\% & 91.4\% & 77.1\% & 42.9\% \\
& Average & 97.4\% & 92.3\% & 82.1\% & 61.5\% \\
\bottomrule
\end{tabular}}
\end{center}
\label{tabel:maintaince}
\end{table}

\begin{findingbox}
Most (Over 90\%) artifacts need continuous maintenance and updating after submission.
The update ratio of earlier artifacts is higher than the newer ones.
\end{findingbox}

\subsection{RQ3: Popularity}
\label{sec:result:popularity}


The popularity of artifacts is closely tied to their impact, especially in the SE field. A more popular artifact has more chances to influence real-world applications and benefit people.
RQ3 focuses on the popularity of artifacts in the field of software engineering. 
We use the number of stars on GitHub as a measure to analyze the popularity of existing SE artifacts. 
Additionally, we study the characteristics of the most popular artifacts.

\subsubsection{Star situation}

\begin{table}[h!]
\caption{The distribution of star numbers of SE artifacts from 2017 to 2022. Each number presents the ratio of artifacts whose star numbers belong to the range in the first column.}
\begin{center}
\resizebox{0.48\textwidth}{!}{
\begin{tabular}{lrrrrrrr}
\toprule
Year&0&1-5&6-10&11-20&21-50&51-100&100+\\
\midrule
2017 & 11.3\% & 42.3\% & 14.1\% & 14.1\% & 7.0\% & 9.9\% & 1.4\% \\
2018 & 13.6\% & 28.2\% & 12.6\% & 13.6\% & 17.5\% & 1.9\% & 12.6\% \\
2019 & 11.7\% & 29.7\% & 18.9\% & 17.1\% & 14.4\% & 1.8\% & 6.3\% \\
2020 & 15.1\% & 39.2\% & 18.1\% & 11.4\% & 10.2\% & 4.2\% & 1.8\% \\
2021 & 12.2\% & 37.6\% & 17.1\% & 14.6\% & 11.2\% & 5.9\% & 1.5\% \\
2022 & 13.5\% & 31.8\% & 19.2\% & 18.8\% & 13.9\% & 0.4\% & 2.4\% \\
\midrule
Average & 13.1\% & 34.6\% & 17.3\% & 15.3\% & 12.5\% & 3.4\% & 3.7\% \\
\bottomrule
\end{tabular}}
\end{center}
\label{tabel:star}
\end{table}

Table~\ref{tabel:star} shows the distribution of the star numbers of the SE artifacts from 2017 to 2022. 

In general, the number of artifact stars is generally small. While 3.7\% of artifacts have over 100 stars, 65.0\% have equal or less than 10 stars. 13.1\% of the artifacts even do not have any stars at all. 
Only eight artifacts have more than 300 stars.
We can observe more clearly that the star numbers concentrate on a low level, indicating the overall low popularity of the SE artifacts.

Furthermore, one might expect a positive correlation between the creation date of an artifact and the number of stars it accumulates. However, our analysis does not reveal a clear relationship between star counts and time. 
To some extent, the 2022 data is indeed more concentrated in the lower number range than the 2017 data, but the data for other years does not exhibit a consistent pattern.

\begin{findingbox}
In general, the popularity of the SE artifacts on GitHub is low, with only 3.7\% of artifacts having over 100 stars and 65.0\% having equal or less than 10 stars. Star numbers and time did not show a clear correlation.
\end{findingbox}

\subsubsection{Characteristics}
The low popularity of existing SE artifacts motivates us to characterize the top-starred artifacts to provide implications for improving future artifacts and their popularity. 

To this end, we consider artifacts with more than 100 stars as top-starred ones. 
By applying this criterion, we identify 33 top-starred artifacts in total.
We compare them with widely adopted criteria in the research community (\cite{acmguidelines}), and discuss how these top-starred artifacts meet the criteria.

According to our analysis, all 33 top-starred artifacts are documented and possess comprehensive documentation, covering the introduction, usage, and examples. 
28 of these top artifacts are accompanied by licenses. 
All the top-starred artifacts are still maintained one year after the publication year of their papers.
\rerevise{
It is worth noting that all the top-starred artifacts include executable code, providing either tools, frameworks, or toolkits.
12 of them have been integrated into large-scale industrial projects, with 4 of them coming from well-known commercial companies like Microsoft and Uber. 
These results indicate that these top-starred artifacts are likely to have a significant impact on real-world applications.
}
Notably, 15 of these artifacts have surpassed 50 forks, while 17 have accumulated more than 50 issues.

As mentioned above, We only focus on the characteristics of these most popular artifacts, and their research objects are beyond the scope of our research.

\begin{findingbox}
Overall, all 33 top-starred artifacts possess comprehensive documentation and are still maintained one year after their publication.
The top-starred artifacts bring a significant impact on real-world applications.
\end{findingbox}

Note that the popularity of artifacts cannot be determined solely based on the number of stars. 
Other metrics, such as issues and forks, could also be taken into consideration. 
To enhance the comprehensiveness of our popularity analysis, we leverage the GitHub API to explore these two metrics for the relevant GitHub artifacts as an additional analytical approach.
We observe that 45\% of GitHub artifacts have no issues, while 80\% have less than ten issues. In terms of fork numbers, nearly 90\% of artifacts have fewer than ten forks. Moreover, the distribution of issues and forks does not exhibit significant differences among artifacts from different years.
Overall, the distribution of these metrics for artifacts closely resembles that of stars. 
Because stars are more prevalent across a larger number of repositories, our analysis primarily focuses on the star number.

\subsection{RQ4: Quality}
\label{sec:result:quality}
The last research question characterizes the quality of artifacts from two aspects: the content of documentation and the circumstances of code smell. 

\subsubsection{Documentation}
\label{sec:result:quality:document}
This section explores the content of the documentation of the artifacts. 

\begin{figure*}[h!]
\begin{center}
\includegraphics[width=0.95\textwidth]{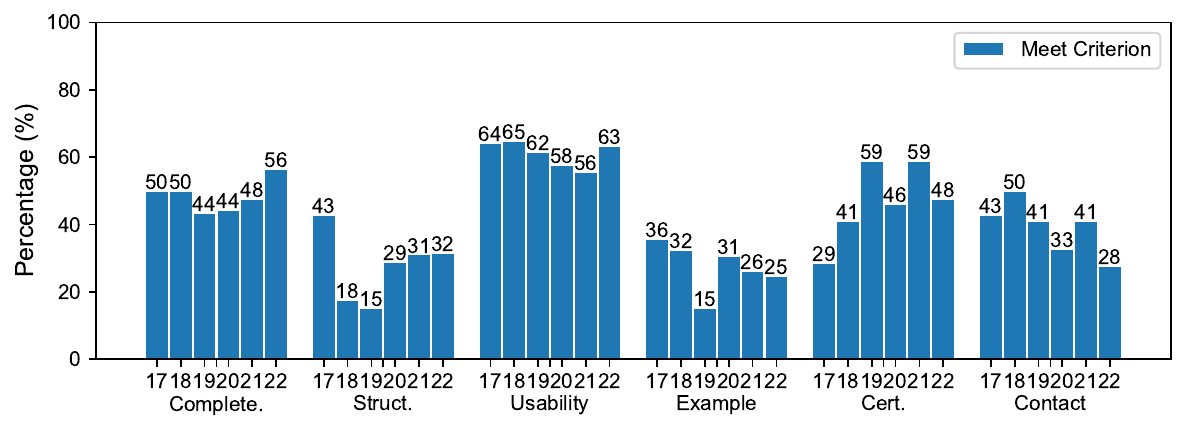}
\end{center}
\vspace{-5mm}
\caption{The documentation situation of ICSE artifacts.}
\label{fig:documentation}
\end{figure*}

The results are shown in Figure~\ref{fig:documentation}.
The completeness of the README went worse, from 50.0\% in 2017 to 43.6\% in 2019. Then it is slowly recovering, finally at 56.4\% in 2022 and with a total average of 49.8\%. 
The ratios of Structure, Usability, and Example are 27.9\%, 60.8\%, and 26.2\% on average. 
The case of Certificate is getting better, from only 28.6\% in 2017 to 47.5\% in 2022. 
The ratio of Contact shows a noticeable decline in 2022, accounting for only 27.7\% in that year.
This may be because it is the newest and not too many issues have appeared.

In addition, we have some interesting findings. The artifacts often meet Completeness and Usability together. 
97.5\% of the artifacts that meet Example at the same time meet Usability altogether. 
The artifacts meeting Structure is often an artifact of a survey, an empirical study, or a data set.

\begin{findingbox}
In general, artifact documentation quality is fair and has not changed much in recent years. Overall, researchers do better in Completeness, Usability, Certificate, and Contact than in the other two criteria, Structure, and Example. There is still room for improvement on all criteria.
\end{findingbox}

\subsubsection{Code Smell}
This section explores the smells of the code researchers write in their artifacts.
We focus on Python and Java code because they are covered by the majority of the artifacts.


\textbf{Python}. We detect the code written in Python by \cite{pylint}. 
Pylint produces code smell results for 268 Python artifacts within a limited time (1 minute), and provides three types of code smells\footnote{While Pylint also provides information in three other categories, including ``Fatal'', ``Error'', and ``Information'', these three are not considered as indicators of a code smell. This is because code smells indicate potentially flawed or suboptimal coding practices instead of bugs or other information.}: 
(1) Convention: the code violates recommended coding conventions; 
(2) Refactor: the code needs effort in refactoring;
(3) Warning: the code contains issues that developers may need to fix, but the issues are not fatal enough to terminate the running of the code.
The detailed content for each type of alert message, explanation, and specific examples can be found in the documentation of \cite{pylintmessageoverview}.

\textbf{Java}.
We use \cite{pmd} to detect the smells in Java code.
PMD successfully runs on all 195 Java artifacts, and we use them as our analysis subjects.
Different from Pylint, PMD provides more diverse alert messages, including
(1) Code style: the code violates a specific coding style;
(2) Documentation: the code needs effort in code documentation;
(3) Design: the code contains design issues;
(4) Best Practices: the code violates generally accepted best practices;
(5) Error Prone: the code contains constructs that are either broken, extremely confusing, or prone to runtime errors;
(6) Performance: the code contains suboptimal code;
(7) Multithreading: the code contains issues when dealing with multiple threads of execution;
(8) Security: the code contains potential security flaws\footnote{Please refer to \url{https://pmd.github.io/latest/pmd_rules_java.html} for more details of each message.}.

\begin{figure}
\begin{minipage}{0.48\linewidth}
  \centerline{\includegraphics[width=\textwidth]{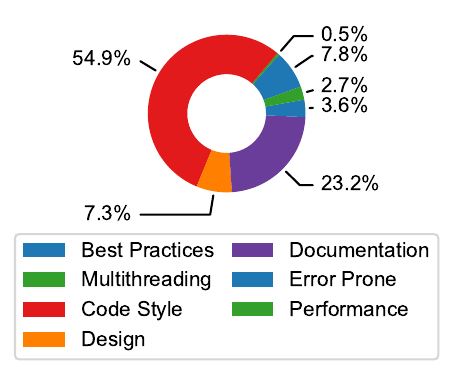}}
  
  \centerline{(a)}
\end{minipage}
\hfill
\begin{minipage}{0.48\linewidth}
  \centerline{\includegraphics[width=\textwidth]{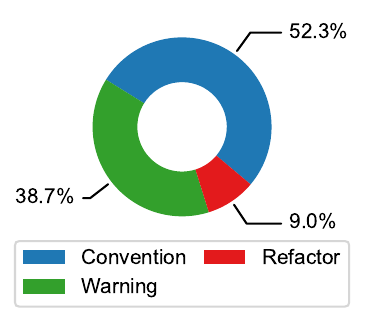}}
  \centerline{(b)}
\end{minipage}
\vfill
\caption{Ratio of different types of Python (a) and Java (b) messages from 2017 to 2022.}
\label{fig:codesmell}
\end{figure}

Figure~\ref{fig:codesmell}.(a) and (b) show the distribution for each type of alert message in Python and Java artifacts.
For Python, the ratio of Convention messages about coding convention violations is the most common, which accounts for 52.3\%. 
There are also 38.7\% warning messages and 9.0\% refactoring alerts. 
The most common smell for Java is code style, similar to convention violations.
These issues may affect the artifacts' readability, maintainability, and usability. It is unsurprising that these issues exist because researchers are often not well-trained developers.
However, poor code styles can make the code hard to understand, maintain, reuse, and even contain potential bugs. While poor code style does not cause major problems in most cases, so many small flaws combined in a single artifact can overwhelm valuable insights in the code, making the paper difficult to understand and reuse. Therefore, code style is also an important part of a paper and should be considered when evaluating papers.

We further dive deep into these alert messages and check their prevalence and specific contents.
Column ``Prevalence'' in Table~\ref{codequalitycount} shows the number of artifacts with the corresponding type of alert messages against the total number of artifacts.
We observe that all the alert messages are surprisingly prevalent in the artifacts.
For example, 254 out of the 268 (98.5\%) artifacts in Python are recommended to refactor their code.  
The average prevalence ratio (the first number divided by the second number in Column ``Prevalence'') of the alert messages is 96.0\% for Python and 98.3\% for Java.

Column ``Name'' and Column ``Explanation'' in Table~\ref{codequalitycount} represent the top three most common alert messages under each message category and their explanation.
We hope that these results could shed light on issues to avoid when researchers wish to improve the quality of their code when preparing artifacts (more details in Section~\ref{sec:Discussion}).

\begin{table*}[h!]
\normalsize
\caption{The top three most common alert messages under each smell category for Python (top three rows) and Java (bottom seven rows).}
\begin{center}
\resizebox{0.99\textwidth}{!}{
\begin{tabular}{lrllr}
\toprule
Type&Prevalence&Name&Explanation&Pct.(\%)\\

\midrule
\multirow{3}*{Convention} & \multirow{3}*{260/268} & invalid-name&Code violates naming rules for variables. & 23.4 \\
 & & line-too-long&Code has a line longer than 100 characters. & 19.9 \\
 & & missing-function-docstring&Code contains a function or method without a docstring. & 19.3 \\
\midrule
\multirow{3}*{Warning} & \multirow{3}*{258/268} & bad-indentation&Code has incorrect indentation. & 58.4 \\
 & &unused-import&Code imports unused modules or variables. & 6.5 \\
 & & redefined-outer-name &Variable in inner scope shares the same name as one in outer scope. & 5.0 \\
\midrule
\multirow{3}*{Refactor} & \multirow{3}*{254/268} & consider-using-with & A resource-allocating assignment may be replaced by a ``with'' block. & 13.1 \\
 & & too-many-locals&Function or method has over 15 local variables. & 10.9 \\
 & & no-else-return&Unnecessary "else" block with a "return" statement. & 9.9 \\

\bottomrule

\toprule

\multirow{3}*{Best Practices} & \multirow{3}*{144/144} & SystemPrintln & Code can use a logger instead of System.out.print references. & 15.8 \\
 &  & LooseCoupling & Code has excessive coupling implementation types. & 8.7 \\
 &  & UnusedAssignment & Code has unused variable assignments. & 8.7 \\
\midrule
\multirow{3}*{Code Style} & \multirow{3}*{144/144} & LocalVariableCouldBeFinal & Local variable assigned only once can be declared final. & 26.5 \\
 &  & MethodArgumentCouldBeFinal & Method argument never reassigned can be declared final. & 22.2 \\
 &  & ShortVariable & Fields, variables, or parameters have very short names. & 11.5 \\
\midrule
\multirow{3}*{Design} & \multirow{3}*{144/144} & LawOfDemeter & Code violates the Law of Demeter. & 29.3 \\
 &  & CyclomaticComplexity & Methods/classes should be broken down into smaller components. & 8.6 \\
 &  & CognitiveComplexity & Methods are too complex. & 6.7 \\
\midrule
\multirow{3}*{Documentation} & \multirow{3}*{144/144} & CommentSize & Comments exceed specified size limits. & 49.8 \\
 &  & CommentRequired & Code needs more comments. & 48.7 \\
 &  & UncommentedEmptyMethodBody & Empty method body without comments. & 1.2 \\
\midrule
\multirow{3}*{Error Prone} & \multirow{3}*{140/144} & AvoidLiteralsInIfCondition & Hard-coded literals used in conditional statements. & 22.1 \\
 &  & AvoidDuplicateLiterals & Duplicate String literals should be declared as constant fields. & 10.7 \\
 &  & DetachedTestCase & Test case method should be detached. & 9.0 \\
\midrule
\multirow{3}*{Multithreading} & \multirow{3}*{134/144} & DoNotUseThreads & Code violates the J2EE specification by using threads. & 34.9 \\
 &  & UseConcurrentHashMap & Code can use the Map designed for multi-threaded access in Java5. & 32.4 \\
 &  & AvoidSynchronizedAtMethodLevel & Method-level synchronization can cause problems with new code. & 21.4 \\
\midrule
\multirow{3}*{Performance} & \multirow{3}*{141/144} & AvoidInstantiatingObjectsInLoops & New objects created within loops should be checked and reused. & 37.8 \\
 &  & RedundantFieldInitializer & Explicit initialization of default values that will be initialized by Java. & 16.7 \\
 &  & ConsecutiveAppendsShouldReuse & Consecutive calls to StringBuffer/StringBuilder appends should be chained. & 8.9 \\

\bottomrule
\end{tabular}}
\label{codequalitycount}
\end{center}
\end{table*}

\begin{findingbox}
Code smells are prevalent in the artifacts.
On average, code smell alert messages appear in 96.0\% of the Python projects and 98.3\% of the Java projects.
The majority of the code smell alert messages are related to code convention violations.
\end{findingbox}

\section{Discussion}
\label{sec:Discussion}

\rerevise{
The growing number of publications with artifacts in the Software Engineering community is a positive indication of progress. 
Artifacts play a crucial role in facilitating the replication and verification of research results, enhancing comprehension of the research, enabling further advancements, and serving as valuable educational resources. 
However, the rise in artifacts presents challenges. 
Firstly, conducting an artifact requires significant additional effort and time from researchers. 
Secondly, preparing functional and well-documented artifacts is technically challenging. 
Lastly, reviewing and evaluating artifacts alongside research papers adds complexity to the peer-review process. 
To address these challenges, the SE community should deepen the research on artifacts and provide comprehensive guidance for artifact preparation, thereby improving the efficiency of managing the entire lifecycle of SE artifacts.
}

In this section, we present our suggestions on artifact preparation and discuss the threats to validity.

\subsection{Suggestion}
Based on our findings, we discuss insights on the status and trends of SE artifacts and provide further suggestions.

\subsubsection{Common Practices}

Regarding \textbf{storage websites} of artifacts, though SE conferences (\cite{icse21policy,icse22policy}) have exploited recommend archiving platforms like Zenodo over version control systems like GitHub, the majority (64.2\% in 2022) of publications choose GitHub for their artifacts.
Because GitHub is easier to maintain and allows for updating the artifacts, we believe GitHub will still be one of the most trending artifact storage websites in the near future.
Additionally, GitHub has a similar link rot rate to Zenodo, which shows that artifacts stored on GitHub can be maintained for a long time.
Besides, most (over 90\%) artifacts have the update requirement after the submission,  indicating that version control systems like GitHub can satisfy the requirement better than archiving platforms like Zenodo. 
Therefore, we suggest that conferences should not blanket exclude version control systems like GitHub, but instead provide guidance on choosing the appropriate platform for specific artifacts.

Moreover, we observe a rapid increase in the use of Zenodo (0.0\% in 2017 to 16.0\% in 2022), which is more suitable for artifacts that do not need to be updated over time, such as datasets. The adoption ratio of Zenodo will likely increase further given its recommendation by conferences as well as its advantages for long-term preservation. We suggest SE researcher choose Zenodo for suitable artifacts.

In terms of \textbf{programming language}, the use of Python has increased from 15.2\% in 2017 to 61.1\% in 2022, surpassing Java. The adoption of Java and C/C++ has decreased over the past six years. We expect this trend to continue going forward.
While quality metrics (e.g. code smells) and detection tools are mature for Java, those for Python are still lacking. 
Given the majority of artifacts now use Python, the SE research community can define coding conventions and specific rules for Python code in artifacts to improve readability and standardization. 
Providing guidance and examination tools for Python artifact code would also enhance quality and reusability.

As for the \textbf{URL location}, we find that 40.9\% artifact URLs are hidden in the middle of the paper and are easily overlooked.
The community has recommended unifying the artifact URL location in the ``Data Availability'' section after the conclusion. 
Although the compliance rate with this rule is still low, there has been a rapid increase in recent years. Unifying artifact URL locations makes it easier to find and access artifacts, increasing their impact. 
We suggest that more publications follow this rule and prominently display artifact URLs in papers, as the increasing importance of artifacts.




\subsubsection{Maintenance}

Regarding artifact availability, the \textbf{link rot} rate increased over the years, with up to around one-third of artifacts inaccessible in 2017. To ensure the long-term accessibility of SE artifacts, the SE community should give additional consideration to artifacts' longevity when reviewing papers, especially for artifacts stored on temporary drives and personal websites that exhibit significantly higher link rot rates than those on GitHub and Zenodo.

For the \textbf{last update time}, most artifacts need continuous updating after submission, highlighting the high updating needs and propensity to become outdated of SE research artifacts. 
Earlier artifacts have a higher update rate. 
We believe that if the community and researchers want to ensure artifacts remain useful and reproducible over time, only one artifact examination before conferences is insufficient.
Given their tendency to become obsolete quickly, regular re-examination is worthwhile for valuable artifacts to keep them up-to-date, effective, and reproducible in the long term.



\subsubsection{Popularity}


Regarding \textbf{star numbers}, most GitHub artifacts receive limited attention, with 65.0\% attracting no more than 10 stars. 
Though artifact star counts are not an accurate measure of popularity, they indicate that most software engineering artifacts lack real-world adoption and impact.
Artifacts may struggle to gain traction because most target niche audiences. 
Even for intended audiences, less than half of respondents report experience reusing artifacts (\cite{ChristopherTimperley2021}).
However, we believe that some prototyped artifacts could have great potential to influence real-world applications and facilitate people's lives if properly disseminated. 
Such artifacts should be viewed not just as paper attachments, but as vital contributions of SE research, advancing the practical implications of software engineering.

We believe that conferences can increase the usefulness and impact of SE artifacts by incorporating \textbf{characteristics} of top-starred artifacts into their evaluation criteria and recommendations, including providing detailed documentation and easy install packages, as well as keeping maintenance.

\subsubsection{Quality}

In terms of \textbf{documentation}, artifact documentation often lacks descriptions of file structure and usage examples, which are vital for most artifacts. 
Describing an artifact's file structure helps users quickly comprehend its composition and working mechanism. 
Usage examples assist users in verifying artifacts that provide tools work properly.
Therefore, We suggest conferences remind researchers to include file structure descriptions and usage examples for suitable artifacts, achieving better documentation quality.

For code quality, we perform \textbf{code smell} tests on Python and Java artifacts and find that over 96\% of artifacts incur alerts. 
Most alerts relate to code convention, not functionality.
As function correctness instead of code convention matters most for code in artifacts, we conclude that the commonly-used metrics for general SE projects, code smells, are unsuitable for artifacts. 
Therefore, we suggest more research on better code quality metrics for artifacts in the SE field, especially for Python.


\subsubsection{Summary}

Based on previous analysis, we summarize suggestions for different stakeholders.

\textbf{For organizers of SE publications.}

(1) Enhance artifact preparation guidance.
Provide comprehensive guidance and detailed assessing criteria for artifact preparation, encompassing aspects such as storage platform selection, document quality enhancement, and subsequent maintenance strategies.

(2) Promote inclusive artifact guidance.
Conferences should avoid blanket exclusions and guide authors in selecting the right platform for their artifacts. 
For static datasets, Zenodo is recommended for archiving, while GitHub is preferred for artifacts involving iterative code updates, with a release version as an archive. 
Conference organizers can provide detailed instructions on different platform features to cater to diverse needs.

(3) Unify the artifact URL in a prominent location. 
As the importance of artifacts becomes increasingly recognized, artifact URLs in papers should be highlighted to facilitate reader access.
Conference organizers can provide a unified recommended position of artifact URLs in the guidance or integrate the URLs into the publication metadata.

(4) Consider artifact longevity during the review process.
Conference organizers should discourage the use of temporary drives or personal websites to host artifacts and consider the long-term validity of artifacts during the review process to minimize the risk of link rot.

(5) Provide high-quality artifact examples.
Offering exemplary artifacts of various types can help authors in producing higher-quality artifacts. For instance, the guidance may present highly acclaimed artifacts from previous years' publications and utilize their characteristics as recommended and evaluative criteria, such as providing easy installation packages, test cases, and comprehensive documents.

(6) Promote real-world adoption and impact.
Conference organizers can encourage researchers to integrate their artifacts into practical application development and promote their potential impact.
We recommend extending the evaluation and recognition of artifacts beyond publications through regular evaluations and giving recognition to those artifacts that demonstrate significant impact.

\textbf{For researchers.}

(1) Emphasize the contribution of artifacts.
To enhance the replication of papers, researchers should strive to provide artifacts for their papers whenever possible and prominently highlight them as a significant contribution to papers.

(2) Choose a suitable storage platform for artifacts.
Researchers should consider factors such as maintenance frequency, long-term storage link availability, sharing options, and data size when selecting a storage platform for artifacts. 
For large static datasets, Zenodo is a recommended choice, while GitHub is suitable for small codebases that require regular updates and maintenance. 
We recommend minimizing the use of personal websites and temporary data drives, which are more prone to link rot.
Moreover, we suggest researchers provide an archived version of the artifacts as a snapshot, regardless of the selected platform.

(3) Enhance artifact quality.
Researchers should improve the quality of artifacts to facilitate the reproducibility of their paper's results by other users. Enhancing artifact quality involves attention to various details, such as comprehensive documentation, clear file structure explanations, simple installation scripts and examples, author contact information, and improved code readability.

(4) Maintain artifacts regularly.
To ensure the long-term usefulness and impact of artifacts, researchers should regularly maintain them. Maintenance tasks include promptly responding to user inquiries, updating dependencies, addressing software versioning issues, and incorporating new features or improvements.

\textbf{For software engineering community.}

(1) Embrace open science practices.
The community should continue supporting and promoting the sharing of open-source artifacts. 
This will foster transparency, reproducibility, and collaboration among researchers.

(2) Promote tools and technologies for artifact management. 
Addressing issues such as link rot and ensuring the enduring availability of artifacts stored outside commonly used platforms can be facilitated by developing tools or leveraging existing technologies like the Wayback Machine and robust links.

(3) Conduct further research on artifact quality metrics. 
The SE community can prioritize research and development of metrics and automated examination approaches specifically tailored to different programming languages and artifact types.
Specifically, the community can offer further guidance and examination tools for Python artifacts, given that Python is the most widely used programming language in current software engineering artifacts.



\subsection{Threats to validity}
\noindent \textbf{Manually check.}
We carry out the empirical study with many manual checks.
Therefore, there are some potential threats to validity.
Multiple checkers are involved in the manual checking process and there is a risk that their judgments may not be consistent.
Besides, the manual checking process could be influenced by biases on the part of the checkers, such as confirmation bias or selection bias. These biases could affect the results of the study and reduce its validity.
Our checkers communicate with each other as much as possible, eliminating subjective differences between different checkers. However, this may still affect.

\noindent \textbf{Selection of storage website.}
We choose artifacts in GitHub to examine their popularity. However, the star number is a  GitHub-specific metric and may not apply to other platforms like Zenodo. 

\rerevise{
\noindent \textbf{Selection of venues.}
We have focused on top-tier software engineering conferences based on the CS Ranking.
Expanding the scope of venue selection and analysis may lead to more valuable findings.
However, the four conferences we selected are well-recognized as top-tier SE venues, which can serve as a good starting point to analyze the state and trends of SE artifacts.
}

\section{Related Work}
\label{sec:Related Work}

In this section, we introduce existing work on community expectations of artifacts. Then, we summarize the work related to the replicability of software engineering.

\subsection{Community Expectations of Artifacts}

Artifact evaluation attracts great attention in software engineering. 
Many previous work propose the recommendations and expectations of artifacts (\cite{hermannCommunityExpectationsArtifacts2020, ChristopherTimperley2021, fanWhatMakesPopular2021, liuReproducibilityReplicabilityDeep2022, di2017software}). Also, some guidelines have been proposed to describe the expected practice of the publications (\cite{fse21guidelines, acmguidelines, icse23policy,vidoni2021software}).

In 2013, 
\cite{krishnamurthi2013artifact} proposed that software and other digital artifacts were amongst the most valuable contributions of computer science but conferences treated these mostly as second-class artifacts. As a result, they highlighted the importance of artifacts and argued for elevating these other artifacts by making them part of the evaluation process for papers. 

\cite{hermannCommunityExpectationsArtifacts2020} provided an overview of the current perception of artifact evaluation and the community expectations toward artifacts. They conducted a survey among the artifact evaluation committees (AEC) members and analysed the purpose and community expectation of the artifact evaluation. In details, they extracted the key information from the response of AEC members and gained the expectations from reviewers and users perspectives.

\cite{ChristopherTimperley2021} conducted an empirical study to understand how artifacts are created, shared, used, and reviewed by mixed methods. They analyzed all research-track papers published at four software engineering venues (ASE, EMSE, FSE, and ICSE) between 2014 and 2018 to determine the prevalence and availability of artifacts, and conducted a survey of 153 authors to identify the challenges of creating, sharing, using, and reviewing artifacts. Based on their finding, they derived several recommendations for different stakeholders in research. Their study is similar to ours to some extend. Our results can better reflect the current status and trends of artifacts because the data we used is from 2017 to 2022. We make many comparison with their findings in our work.

We also compare our findings with the guidelines of 
\cite{fse21guidelines} and 
\cite{icse23policy}, and the 
\cite{acmguidelines} guidelines, which describe the expected characteristics of artifacts of publications in the above sections.

\subsection{Replicability in Software Engineering}

Replicability has become an important concern in software engineering research (\cite{flittner2018survey, saucez2019evaluating}), and ongoing work focuses on the quality, reliability, and trustworthiness of software engineering research results through the development and application of effective replicability techniques.

\citeauthor{liuReproducibilityReplicabilityDeep2022} 
conducted a literature review on 147 DL studies published in 20 SE venues and 20 AI venues to investigate these issues and re-ran four representative DL models in SE to investigate important factors that may strongly affect the reproducibility and replicability of a study. They concluded that it is urgent for the SE community to provide a long-lasting link to a high-quality reproduction package, enhance DL-based solution stability and convergence, and avoid performance sensitivity on different sampled data, which is proved by our work another time.

\citeauthor{8951162} aim to identify tools that maximize reproducibility in software engineering experiments and how they are applied and performed a systematic mapping study and complementary strategies to analyze replication from communication concern, knowledge management concern, and motivation concern. They concluded that reproducibility is mostly relegated to internal replication, at which time and costs can be assumed within research groups and a focus on new alternatives should be considered to broaden replication.

Our work confirms these results and puts forward recommendations on how to better prepare artifacts, which will help improve replicability.

\section{Conclusion and Future Work}
\label{sec:Conclusion}

We conduct an empirical study to understand the current status and trends of research artifacts in SE publications.
We conduct a dataset with 2,196 papers and corresponding 1,487 artifacts for analysis.
Based on the dataset, We investigate the common practices, maintenance, popularity, and quality. 
We observe an increase in the number of papers providing artifacts in SE venues.
The most commonly used storage platform is still GitHub, while the employment of Zenodo rapidly grows.
Over 90\% of artifacts would be updated after the submission.
At the same time, 65.0\% of artifacts have less than 10 stars and the code smell alert messages appear in 96.0\% of the Python projects and 98.3\% of the Java projects.
Moreover, we provide the discussion on the overall status of SE artifacts and offer suggestions to different stakeholders.
We hope our data and empirical results can support further study on the research artifacts of the SE community. 

\noindent \textbf{Future Work.}
To capture the latest status and trends in the SE community, we study publications of four top-tier SE venues according to the CS Ranking. 
However, we could uncover more interesting findings if we expand our methodology and include a broader range of SE venues, including journals and more popular SE conferences with various requirements for artifacts.

Note that the annotation work is labor-intensive, as it entails manually searching for the artifact URLs for each paper and manually accessing, downloading, and analyzing them. As a starting point, we have developed a website, CS-Artifacts\footnote{http://ra.bdware.cn}, to showcase our annotated papers and their corresponding artifact data. We aim to gradually expand the coverage of venues on this website in the future.

\section{Data Availability}
\label{Data Availability}

We make our data, scripts, and results publicly available at \url{https://github.com/morgen52/SE-artifact}.

\section*{Acknowledgment}
This work was supported by the National Key Research and Development Program of China under the grant number 2021YFF0901100, and Center for Data Space Technology and System, Peking University.

\bibliographystyle{cas-model2-names}

\bibliography{paper_artifact}



\end{document}